\documentclass[letterpaper]{article} 
\usepackage{aaai2026}  
\usepackage{times}  
\usepackage{helvet}  
\usepackage{courier}  
\usepackage[hyphens]{url}  
\usepackage{graphicx} 
\usepackage{natbib}  
\usepackage{caption} 
\usepackage{algorithm}
\usepackage{algorithmic}
\usepackage{marvosym}

\usepackage{newfloat}
\usepackage{listings}
\DeclareCaptionStyle{ruled}{labelfont=normalfont,labelsep=colon,strut=off} 
\floatstyle{ruled}
\newfloat{listing}{tb}{lst}{}
\floatname{listing}{Listing}
\usepackage{booktabs}
\usepackage{amssymb}
\usepackage{amsmath}
\usepackage{multicol}
\usepackage{multirow}

\title{GuideGen: A Text-Guided Framework for Paired Full-torso Anatomy and CT Volume Generation}
\author {
    Linrui Dai\textsuperscript{\rm 1,2}\equalcontrib\thanks{This work was done when the author was a master student at Shanghai Jiao Tong University.},
    Rongzhao Zhang\textsuperscript{\rm 3}\equalcontrib,
    Yongrui Yu\textsuperscript{\rm 1},
    Xiaofan Zhang\textsuperscript{\rm 1}\thanks{Corresponding author.}
}
\affiliations {
    \textsuperscript{\rm 1}Shanghai Jiao Tong University\\
    \textsuperscript{\rm 2}The University of Tokyo\\
    \textsuperscript{\rm 3}Shanghai Artificial Intelligence Laboratory\\
    
    o-o111@g.ecc.u-tokyo.ac.jp, zhangrongzhao@pjlab.org.cn, \{yuyongrui, xiaofan.zhang\}@sjtu.edu.cn
}

\begin{document}

\maketitle
\begin{abstract}
The recently emerging conditional diffusion models seem promising for mitigating the labor and expenses in building large 3D medical imaging datasets. However, previous studies on 3D CT generation primarily focus on specific organs characterized by a local structure and fixed contrast and have yet to fully capitalize on the benefits of both semantic and textual conditions. In this paper, we present GuideGen, a controllable framework based on easily-acquired text prompts to generate anatomical masks and corresponding CT volumes for the entire torso—from chest to pelvis. Our approach includes three core components: a text-conditional semantic synthesizer for creating realistic full-torso anatomies; an anatomy-aware high-dynamic-range (HDR) autoencoder for high-fidelity feature extraction across varying intensity levels; and a latent feature generator that ensures alignment between CT images, anatomical semantics and input prompts. Combined, these components enable data synthesis for segmentation tasks from only textual instructions. To train and evaluate GuideGen, we compile a multi-modality cancer imaging dataset with paired CT and clinical descriptions from 12 public TCIA datasets and one private real-world dataset. Comprehensive evaluations across generation quality, cross-modality alignment, and data usability on multi-organ and tumor segmentation tasks demonstrate GuideGen's superiority over existing CT generation methods. Relevant materials are available at https://github.com/OvO1111/GuideGen. 
\end{abstract}

\section{Introduction}
\begin{figure*}[ht]
    \centering
    \includegraphics[width=0.9\linewidth]{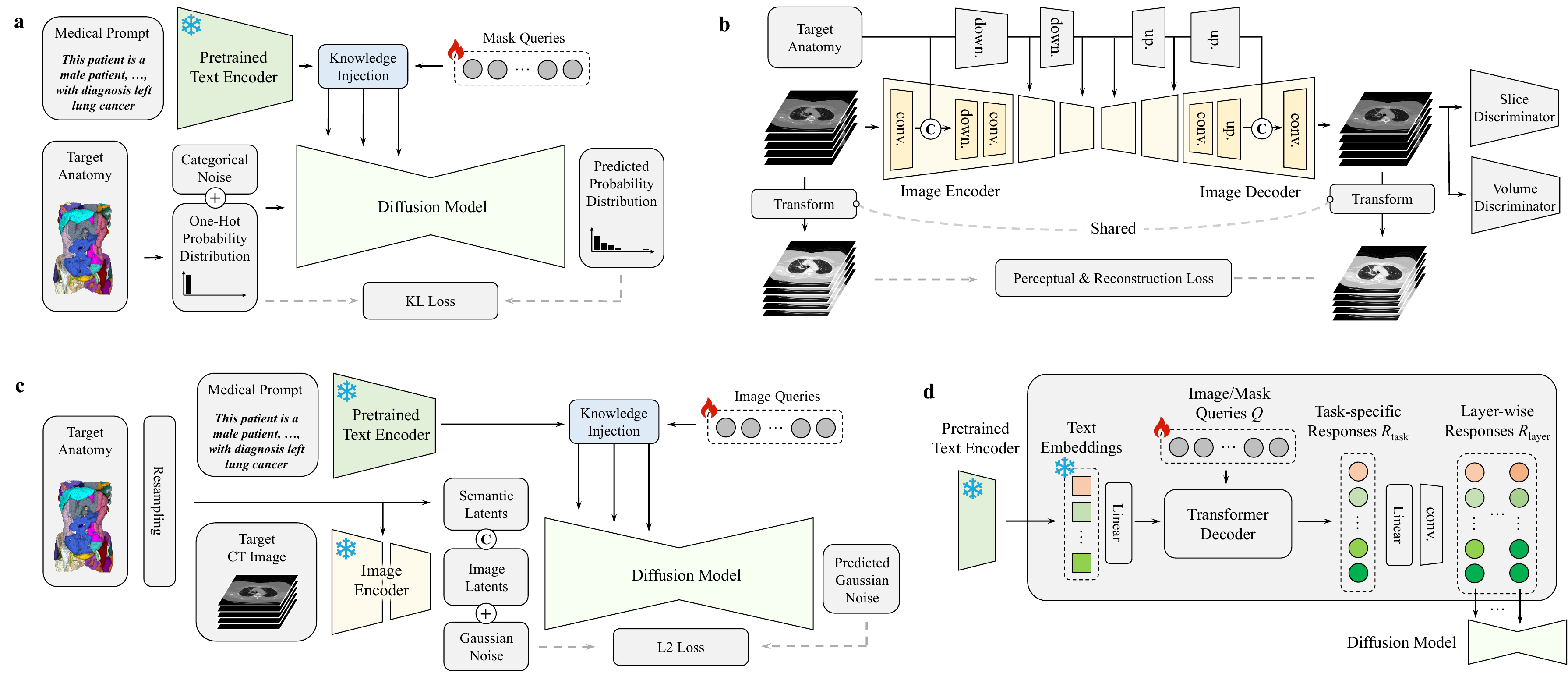}
    \caption{Overview of GuideGen's training pipeline: \textbf{(a)} Firstly, GuideGen learns to generate discrete semantic volumes that conforms to spatial features designated in the medical prompt (See Sec.3.1); \textbf{(b)} Secondly, GuideGen deploys a pyramidal autoencoding scheme to incorporate mask knowledge and reconstruct fine CT details with a high dynamic range (See Sec.3.2); \textbf{(c)} Finally, GuideGen combines the semantic latents derived from (a), image latents extracted in (b) and textual latents from the medical prompt to synthesize full-torso CT images (See Sec.3.3); \textbf{(d)} The internal structure of our knowledge injection module for extracting task-specific features from a structured input used in (a) and (c).
    }
    \label{fig:pipeline}
\end{figure*}

The acquisition of a large quantity of medical images and their corresponding labels has always been critical for modern medical image analysis. However, due to data privacy issues and laborious annotation work, such datasets are often unavailable publicly, undermining the performance of subsequent tasks~\cite{dai2023efficient}. Recently trending conditional generative methods offer a promising solution to these problems: By eliminating privacy concerns and human interventions, they can generate vivid samples to an arbitrary scale, which provide significant boost to the performance of downstream applications like image segmentation and classification~\cite{huang2024label,hashmi2024xreal}.  

Current conditional generative approaches can be categorized based on the nature of the conditions they incorporate. Semantic conditions, like organ and tumor maps, can be harnessed to generate images that adhere to specific locality constraints~\cite{yao2021label,yang2023diffusion,hu2023label,han2023medgen3d,guo2024maisi}. Textual conditions, on the other hand, have the advantage of enhanced generation diversity and reduced human effort at sample synthesis thanks to the versatility of natural language~\cite{chambon2022roentgen,balaji2022ediffi,cho2024medisyn,huang2024chest,guo2025text2ct}. However, current advancements that adopts conditional pipelines seldom leverage the merits from both categories. Consequently, semantic-guided models typically exhibit limited sample diversity and depend on comprehensive anatomy masks that are costly to curate. Conversely, text-guided models, though much more flexible, struggle to fully capture exact spatial relationships among anatomical structures. Realizing this disadvantage, Kim \textit{et al.}~\cite{kim2024controllable} proposes to use a text-guided controllable generation pipeline that generates CT images not only conforms to a general semantics, but also to a textual guidance on image specifics. However, it requires a paired input of semantic and textual guidance as input, which greatly limits its applicability. MedSyn~\cite{xu2024medsyn} alleviates the need of paired input at inference by setting a null semantic input and recovering CT from a learned joint distribution. However, it remains unclear whether its chest-only pipeline can broadcast to the entire torso with more complex anatomical structures. Also, current text-guided generative pipelines largely focus on straightforward usages in classification tasks, and their potential for downstream segmentation remains underexplored, which is both critical in clinical treatment planning and accurate diagnosis~\cite{xu2024advances}.

To this end, we propose GuideGen, a novel framework that synthesizes full-torso anatomies and CT images based on medical strctured inputs. Naturally, GuideGen can synthesize paired samples that provides for dataset construction on downstream segmentation tasks. It is also more user-friendly than mask-based competitors as it necessitates only a textual input, empowering researchers to synthesize full-torso CT dataset with minor effort. GuideGen first offers a text-conditional semantic synthesizer based on a categorical diffusion model to generate discrete label indices, unlike current practices~\cite{han2023medgen3d} that suffer from an ambiguity problem (See Sec.~\ref{sec:TCSS}). We also develop a knowledge injection module to help translate implicit information in medical prompts. Secondly, we train an anatomy-guided autoencoder that extracts comprehensive anatomical features from multiple contrast levels, avoiding the detail degradation by truncating intensities to a specific range that favors a region of interest~\cite{chen2024towards,yu2024ct} while preserving small organ and tumors described in the semantic condition. Finally, we utilize a diffusion latent generator that operates on the combined space of generated semantics and input textual features to recover valid image latents. 

We train and validate GuideGen on an assembly of 12 public tumor datasets and one in-house colorectal cancer dataset, which contain the multi-modal information of CT images and medical descriptions. To testify GuideGen's downstream usability for segmentation tasks, we further include 2 multi-organ segmentation datasets and 3 tumor segmentation datasets. Our framework achieves state-of-the-art results in sample quality, conditional consistency as well as downstream usability on segmentation tasks across multiple datasets, all with the help of single textual inputs.

\section{Related Works}
\noindent\textbf{Generative frameworks}, such as Energy-Based Models~\cite{du2019implicit,guo2023egc}, Generative Adversarial Networks~\cite{de2021impact,zhou2023gan}, Normalizing Flows~\cite{hajij2022normalizing,jeevan2024normalizing}, Variational Autoencoders~\cite{kingma2022autoencoding}, and Diffusion Models~\cite{yoon2023sadm,hung2023med,iglesias2024generation} encouraged myriad researches on image generation due to their high authenticity and variability. Among these studies, the most advantageous feature that has emerged is arguably the capacity to steer the generation process via user-defined conditions~\cite{po2024state}. This capability allows for a high degree of precision and customization in creating images, responding directly to the specific requirements set by the user, while providing a way for the generated images to be used for dataset augmentation purposes~\cite{fang2024data,islam2024diffusemix}. In this work, we extend the ability of current diffusion models for a text-guided anatomy and CT synthesis, which mitigates data scarcity experienced in the field of medical image analysis.

\noindent\textbf{Conditional Medical CT Synthesis}, on which previous studies delivered convincing gains on selected downstream tasks~\cite{hashmi2024xreal,kazerouni2023diffusion,colleoni2024guided,konz2024anatomically,hu2023label,guo2024maisi}. Despite their pioneering efforts, their work mostly rely on external semantic guidance at inference, which is not costly to attain at clinical practices. On the other hand, text-based generation pipelines~\cite{pinaya2022brain,hamamci2025generatect,guo2025text2ct} usually supplies too little information to yield high-quality volumes and can only benefit downstream classification tasks. Although recent work has begun to combine the merits from both worlds~\cite{kim2024controllable,xu2024medsyn}, they require paired inputs at inference or generate only regional patches and cannot provide for downstream segmentation tasks. While a step forward, this indicates room for further development to achieve a more automated and generalized solution. This paper intends to provide a text-guided full-torso generative framework that aids to a wider range of downstream tasks.

\section{Method}
As illustrated in Fig.~\ref{fig:pipeline}, GuideGen separates the generative process into three stages to generate full-torso anatomy and CTs for downstream tasks with text-only inputs. We will elaborate on our designs in the following sections.

\subsection{Text-conditional Semantic Synthesizer}
\label{sec:TCSS}
\noindent\textbf{Ambiguity-reducing Categorical Modeling} Current generative frameworks that uses dedicated semantic synthesizers~\cite{han2023medgen3d,chu2024anatomic} unintentionally introduce ambiguities as they struggle to capture the sharp transitions in labels near semantic boundaries due to an inaccurate data modeling. 
Contrary to these methods, our Text-Conditional Semantics Synthesizer (TCSS) provides a valid solution to reduce ambiguity by building upon a categorical diffusion model~\cite{hoogeboom2021argmax,zbinden2023stochastic} which directly assumes a discrete formulation. As shown in Fig.~\ref{fig:pipeline}(a), at its core, we train a categorical diffusion $p_\theta$ parameterized by $\theta$ that models the underlying distribution of mask volumes $\mathbf{m}$. At the first timestep, the diffusion variable $\mathbf{x}_0=\mathbf{m}\in\{1,\cdots,N\}^{H\times W\times D}$, where $H$, $W$, $D$ separately denotes mask height, width and depth. In the forward process, $\mathbf{x}_0$ is gradually transformed to a categorical noise $\mathbf{x}_T\sim\mathcal{C}_N(\mathbf{x}_T;\mathbf{1}/N)$ in $T$ timesteps under a noise schedule $\beta_{1:T}$, where $\mathcal{C}_N$ denotes a categorical distribution of $N$ categories (semantic classes). The forward probability transition can be described by
\begin{equation}
    q(\mathbf{x}_t^i|\mathbf{x}^i_{t-1})=\mathcal{C}_N(\mathbf{x}_t^i;(1-\beta_t)\mathbf{e}(\mathbf{x}^i_{t-1})+\beta_t\cdot\frac{\mathbf{1}}{N})\in[0,1]^N,
\end{equation}
where the superscript $i$ is the voxel index and $\mathbf{e}(\cdot)$ is a function that returns one-hot probability vectors from a categorical input. For compactness and ease of reading, from now on we will consider the diffusion process on an image with only one voxel to lose the superscript $i$, \textit{i.e.} $H=W=D=1$. Supposing voxels are independent, the formula derived below can be safely extended to images of any size. 
From the properties of Markov chains~\cite{hoogeboom2021argmax}, the forward process of $p_\theta$ can be described as
\begin{equation}
q(\mathbf{x}_t|\mathbf{x}_0)=\mathcal{C}_N(\mathbf{x}_t;\bar{\alpha}_t\mathbf{e}(\mathbf{x}_0)+(1-\bar{\alpha}_t)\cdot\frac{\mathbf{1}}{N})\in[0,1]^N,
\end{equation}
where $\bar{\alpha}_t=\prod_{\tau=1}^t(1-\beta_\tau)$.

\noindent\textbf{Knowledge Injection} To provide an interface for more fine-grained and versatile control via structured prompts, we extract relevant information in the medical prompt for semantic placement. Conventionally, it is widely adopted to use a dedicated text encoder $\mathcal{E}_{T}$ pretrained on medical knowledge~\cite{sun2021ernie} to map the original prompt $\mathbf{p}$ onto a latent space and perform cross-attention with the generation backbone~\cite{chambon2022roentgen,xu2024medsyn,hamamci2025generatect}. However, this alone is often insufficient to steer the model's focus to the most pertinent descriptions in input medical prompts. For instance, the model may fail to prioritize critical information like tumor location over less relevant descriptors, such as race and gender, during semantic synthesis. Recognizing this, we opt to use a separate module to extract task-specific features from encoded latents. 

Specifically, as shown in Fig.~\ref{fig:pipeline}(d), knowledge injection works by allowing learnable task-specific generation queries $Q$ to interact with a series of transformer decoder blocks to retrieve relevant responses $R_{\rm task}\in\mathbb{R}^{N\times C}$ for mask or image generative tasks from encoded medical prompts $\mathcal{E}_T(\mathbf{p})$. In addition, since different layers in the generative backbone perceive different levels of semantic information, we derive layer-wise responses $R_{\rm layer}\in\mathbb{R}^{N\times(C\times L)}$ focusing on either global anatomies or local structures from $R_{\rm task}$. $N$, $C$, $L$ separately denote the number of query tokens, latent dimension and network depth of $p_\theta$. This layer-wise guidance $R_{\rm layer}$ is then injected into the diffusion backbone $p_\theta$: For each intermediate layer $l$ in the diffusion backbone $p_\theta$, denoting its textual guidance as $R_{\rm layer}^l$, the generative latents $\mathbf{z}_l$ at layer $l$ is computed by
\begin{gather}
\mathcal{Q}=W^q(\mathbf{z}_{l-1}),\mathcal{K}=W^k(R_{\rm layer}^l),\mathcal{V}=W^v(R_{\rm layer}^l)\\
    \mathbf{z}_{l} = \text{MLP}(\text{LayerNorm}(\text{Softmax}(\mathcal{Q}\mathcal{K}^\top/\sqrt{d})\mathcal{V})),
\label{eq:transformer}
\end{gather}
where $W^q$, $W^k$, and $W^v$ are learnable projections and $d$ is the context dimension.

\noindent\textbf{Training Objective} We train $p_\theta:[0,1]^N\rightarrow[0,1]^N$ by minimizing the Kullback-Leiber divergence between posterior and generator distributions across the diffusion process as in~\cite{ho2020denoising}. Denoting the posterior $q(\mathbf{x}_{t-1}|\mathbf{x}_t,\mathbf{x}_0)=\frac{q(\mathbf{x}_t|\mathbf{x}_{t-1})q(\mathbf{x}_{t-1}|\mathbf{x}_{0})}{q(\mathbf{x}_t|\mathbf{x}_{0})}\in[0,1]^N$ as $\boldsymbol{\pi}_t(\mathbf{x}_0)$, the overall training objective is
\begin{equation}
\mathcal{L}=\mathbb{E}_{t,\mathbf{x}_0}\left[\mathcal{D}_{\rm KL}\left(\mathcal{C}_N(\mathbf{x}_{t-1};\boldsymbol{\pi}_t(\mathbf{x}_0))||\mathcal{C}_N(\hat{\mathbf{x}}_{t-1};p_\theta)\right)\right],
\end{equation}
with the hat notation distinguishing generated variables.

To avoid training instability, we make a similar reparameterization as in DDPM~\cite{ho2020denoising} for enhanced performance by training a generator model on the same space as the input $\mathbf{x}_0$. Instead of training a model $p_\theta(\mathbf{x}_t,\mathbf{p})$ that models on the distribution $\mathcal{C}_N(\hat{\mathbf{x}}_{t-1};p_\theta)$, we will train and draw samples from its reparameterized version, denoted as $f_\theta=\mathcal{C}_N(\hat{\mathbf{x}}_{0};f_\theta(\mathbf{x}_t,\mathbf{p}))$. The training loss $\mathcal{L}$ needs to be modified accordingly: From the total probability formula, denoting $\mathbf{A}=(\boldsymbol{\pi}_t(1),\cdots,\boldsymbol{\pi}_t(N))$, we have $p_\theta=\mathbf{A}f_\theta$. The voxel-wise loss for TCSS can be written as 

\begin{equation}
    \mathcal{L}_{\rm 1}=
    \mathbb{E}_{t,\mathbf{x}_0}\left[\mathcal{D}_{\rm KL}\left(\mathcal{C}_N(\mathbf{x}_{t-1};\boldsymbol{\pi}_t(\mathbf{x}_0))||\mathcal{C}_N(\hat{\mathbf{x}}_{t-1};\mathbf{A}f_\theta)\right)\right].
\end{equation}
Under real settings where an image actually contains multiple voxels, $\mathcal{L}_{\rm 1}$ is averaged over each individual voxel.

At inference, TCSS samples from the distribution $f_\theta$ and computes the next denoising step according to a sampling scheme $S$ on the generator distribution $S(\mathbf{A}f_\theta(\hat{\mathbf{x}}_t,\mathbf{p}))$. Similar to~\cite{zbinden2023stochastic}, we sample by probability for timesteps $2\sim T$ and take the index with max probability on the final denoising step to add stability to the generation process. Finally, generated masks can be retrieved via $\hat{\mathbf{m}}=\rm{argmax}(\hat{\mathbf{x}}_0)$ along the channel dimension.

\subsection{Anatomy-aware HDR Autoencoder}
\label{sec:CAE}
\noindent\textbf{Anatomy Preservation} Similar to previous works~\cite{cho2024medisyn,hamamci2025generatect}, we use a pair of image autoencoder to extract high-level texture details from CT volumes while lowering memory demands to incorporate larger CT volumes. However, current practices that directly apply off-the-shelf autoencoder architectures could undermine the preservation of spatial relations among anatomies of different sizes. For example, a small tumor visible in a higher resolution input may not be as readily recognized by the decoder half or the subsequent diffusion model in the latent space of the autoencoder. Therefore, we propose to use semantic masks to constrain the autoencoding process and mitigate this undesirable effect. As illustrated in Fig.~\ref{fig:pipeline}(b), we follow a pyramidal approach to resample the semantic information to the resolution of layer-wise latents in the autoencoder and concatenate them in each layer. This helps the model focus on semantic details at encoding and reconstruct semantically accurate images in the decoding process.

\noindent\textbf{HDR Accommodation} Secondly, unlike concurrent works that truncate input CT volumes to a limited intensity range~\cite{yu2024ct,chen2024towards,zhuang2023semantic}, we aim to accommodate the entire dynamic range of intensities in full-torso CTs to incorporate details for each organ. Similar to physicians examining different anatomies using different intensity windows, we devise an intensity transformation module $h$ and constrain the reconstruction process under the full spectrum of intensity ranges. Formally, 
\begin{equation}
    h(\mathbf{x})=k\max\left\{\min\left\{\frac{\mathbf{x}-w_c+w_r}{2w_r},1\right\},0\right\}+b,
\end{equation}
where $w_c$ and $w_r$ are separately known as the window center and window radius. By definition, $h$ truncates the intensity range of $\mathbf{x}$ to $[w_c-w_r,w_c+w_r]$ and rescales them to $[b,k+b]$. At each training step, we randomly sample $w_c$ and $w_r$ from the intensity range of $\mathbf{x}$ and use learnable coefficients $k$ and $b$ to map the truncated result back to the input space. By denoting the encoder and decoder halves of our autoencoder as $\mathcal{E}_I$ and $\mathcal{D}_I$, the $l_1$ reconstruction loss on an input volume $\mathbf{v}$ can be defined as
\begin{equation}
\mathcal{L}_{\rm rec}=||h(\mathcal{D}_I(\mathcal{E}_I(\mathbf{v})))-h(\mathbf{v})||.
\end{equation}

Aside from $\mathcal{L}_{\rm rec}$, to reduce noise in the reconstructed images, we incorporate adversarial components into our autoencoders. Specifically, we include a frame discriminator $\mathbf{D}_f$ that randomly chooses several planar slices from the CT volume and tries to distinguish between authentic slices and synthetic ones, and a volume discriminator $\mathbf{D}_v$, that does the same job from a volumetric viewpoint. These discriminators, separately focusing on anatomical details within CT slices and the general structure of CT volumes, are not subject to intensity transformations $h$ to avoid training instability and preserve the range of intensities for each organ. Furthermore, we use perceptual loss based on a VGG-16 network~\cite{zhang2018perceptual,simonyan2014very} $\mathcal{L}_{\rm perc}$ for reducing artifacts as in~\cite{chen2024towards}. Overall, the loss function for an input CT volume $\mathbf{v}$ is
\begin{equation}
\begin{split}
\mathcal{L}_{\rm 2}=&\mathcal{L}_{\rm rec}(h(\mathcal{D}_I(\mathcal{E}_I(\mathbf{v}))),h(\mathbf{v}))+\\&\mathcal{L}_{\rm perc}(h(\mathcal{D}_I(\mathcal{E}_I(\mathbf{v}))),h(\mathbf{v}))+\\&\mathcal{L}_{\rm disc}(\mathcal{D}_I(\mathcal{E}_I(\mathbf{v}))[i],\mathbf{v}[i],\mathbf{D}_f)+\\&\mathcal{L}_{\rm disc}(\mathcal{D}_I(\mathcal{E}_I(\mathbf{v})),\mathbf{v},\mathbf{D}_v),
\end{split}
\end{equation}
where $\mathcal{L}_{\rm disc}(\mathbf{x}',\mathbf{x},\mathbf{D})=\log \mathbf{D}(\mathbf{x})+\log(1-\mathbf{D}(\mathbf{x}'))$.

\subsection{Latent-guided Feature Generator}
\label{sec:LFG}
In the final stage, we use a diffusion model to approximate the distribution of image latents $\mathbf{z}_0$ encoded by our autoencoder based on the textual and semantic guidance from input medical prompt and TCSS. In the forward process, $\mathbf{z}_0$ is gradually converted to Gaussian noise $\mathbf{z}_T\sim\mathcal{N}(0,\mathbf{I})$ through $T$ timesteps following a noise schedule $\beta'_{1:T}$
\begin{equation}
    q(\mathbf{z}_t|\mathbf{z}_{t-1})=\mathcal{N}(\mathbf{z}_t;\sqrt{1-\beta'_t}\mathbf{z}_{t-1},\beta'_t\mathbf{I}).
\end{equation}

As illustrated in Fig.~\ref{fig:pipeline}(c), We follow a similar way as in our TCSS to use knowledge injection and cross-attention layers to inject textual information into our diffusion backbone and concatenate resampled semantic to the diffusion variable $\mathbf{z}_t$ at each timestep (Fig.~\ref{fig:pipeline}(c)). The generative objective is constructed as in~\cite{ho2020denoising}:
\begin{equation}
    \mathcal{L}_{\rm 3}=\mathbb{E}_{t,\boldsymbol{\epsilon}\sim\mathcal{N}(0,\mathbf{I})}\left[||\boldsymbol{\epsilon}-f_\varphi(\mathbf{z}_t;Resample(\hat{\mathbf{m}}),\mathbf{p})||_2^2\right],
\end{equation}
where $f_\varphi$ is the 3D U-Net~\cite{cciccek20163d} diffusion backbone parameterized by $\varphi$ and controlled by $\hat{\mathbf{m}}$ after being resampled to a desired output resolution as well as textual conditions $\mathbf{p}$. At inference time, $f_\varphi$ gradually recovers an image latent $\hat{\mathbf{z}}_0$ by reverse sampling from random noise, which is then decoded into a valid CT image by the decoder $\mathcal{D}_I$ described in Sec.~\ref{sec:CAE}. The generated sample pair $\{Resample(\hat{\mathbf{m}}),\mathcal{D}_I(\hat{\mathbf{z}}_0)\}$ can then be used for constructing or augmenting segmentation dataset.

\section{Experiments}
\begin{figure*}[!t]
    \centering
    \includegraphics[width=\linewidth]{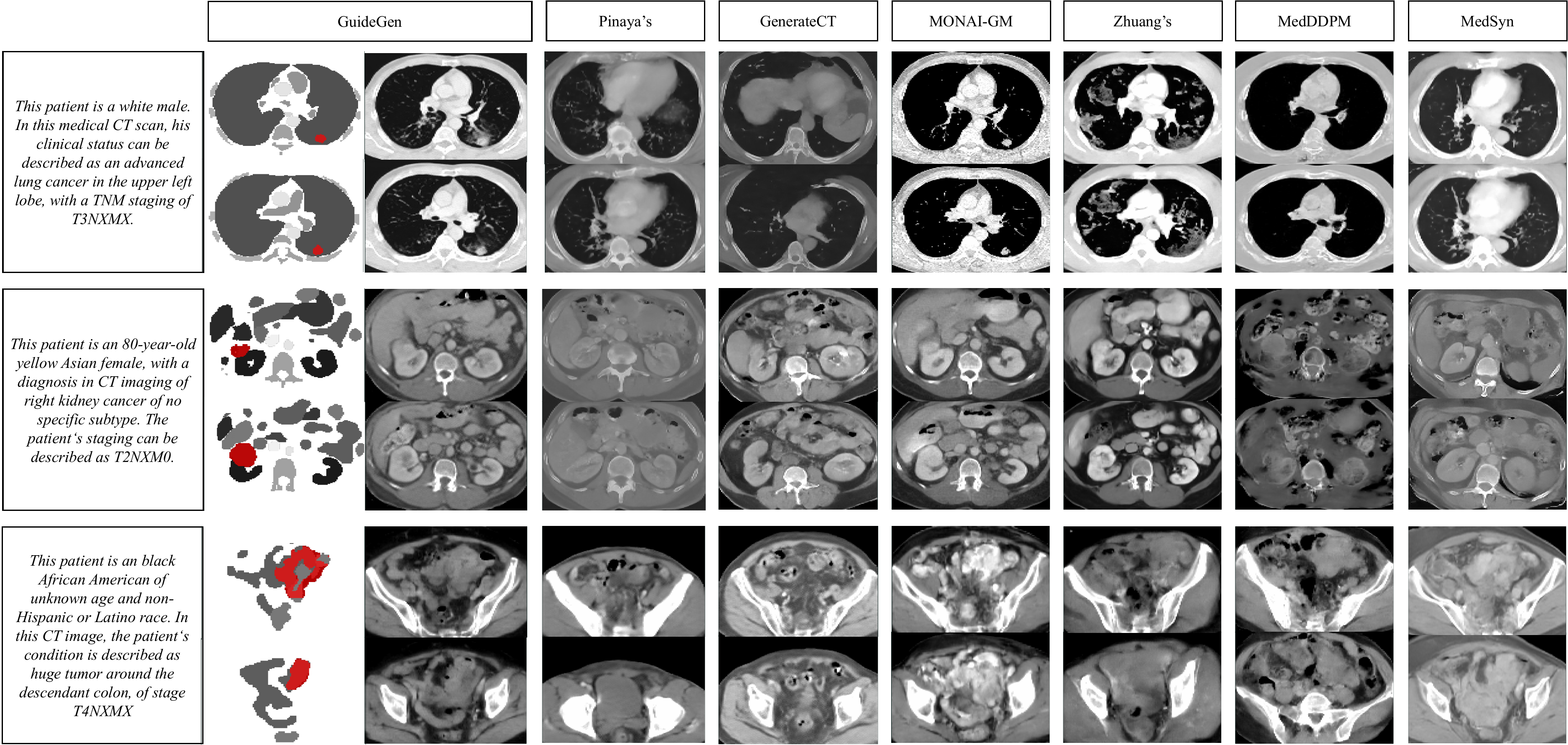}
    \caption{Qualitative results of different generation methods conditioned on the same textual prompts. Mask inputs to baseline models are generated with GuideGen with tumor semantics masked in red (the first column). Two slices are shown per case. For better visualization, we use a CT intensity window of [-975,-225]HU for displaying the chest region (rows 1-2), and [-50,150]HU for the abdominal region (rows 3-6)~\cite{seeram2015computed}. See our project page for more qualitative results.}
    \label{fig:results}
\end{figure*}
\label{sec:exp}
\paragraph*{Dataset Construction}
(1) \textbf{Training}: We trained GuideGen on a compiled CT dataset from 12 public TCIA sources~\cite{clark2013cancer} and a private dataset (RJ), which supplements scarce public colorectal cancer data. Text prompts were generated by a private medical LLM that converted structured records (TCIA) and reports (RJ) into free-text descriptions using the template: ``\textit{The patient is \{demographics group\}. In this imaging, the patient's condition is described as \{clinical information\}}'' For efficiency, ground-truth (GT) segmentation maps were pseudo-labeled using pre-trained networks~\cite{wasserthal2023totalsegmentator,isensee2021nnu}. This dataset was split into 4534 training and 1179 validation cases. (2) \textbf{Inference}: For downstream task evaluation, we use BTCV~\cite{landman2015miccai} and AMOS22~\cite{ji2022amos} for multi-organ segmentation as well as lung tumors (LU), colon cancer (CO) in MSD dataset~\cite{antonelli2022medical} and KiTS21~\cite{heller2023kits21} for tumor segmentations. All other experiments used the validation split of our collected TCIA/RJ data.

\noindent\textbf{Evaluation Metrics}
We use Perceptual Image Patch Similarity (LPIPS), Fr\'{e}chet Inception Distance (FID, averaged on all slices from 3 axes) and Fr\'{e}chet Video Inception Distance (FVD, axial slices as frames) to evaluate the generation quality~\cite{park2023learning,yu2024ct,lei2025lesiondiffusion}. Additionally, we report the Dice Similarity Coefficient (DSC) to evaluate image-mask consistency and the accuracy of several dedicated classifiers to measure alignment between generated CT and features described in the input prompt. Downstream segmentation performance is evaluated through DSC and 95\% Hausdorff Distance (HD$_{95}$).

\noindent\textbf{Implementation Details}
All generative trainings are performed under a constant learning rate of $2\times 10^{-5}$ with a batch size of 1 per GPU, a mask size of $128^3$ and an image size of $256^3$. Training processes use AdamW~\cite{loshchilov2017decoupled} optimizers with a momentum of 0.99 and a weight decay of $1\times10^{-5}$. We implement each diffusion model with a cosine noise schedule and 1000 timesteps using PyTorch~\cite{paszke2019pytorch}. GuideGen is trained and tested on 4 NVIDIA 4090 GPUs with max VRAM usages of 23.2GB and 8.9GB separately for training and inference.

\subsection{Main Results}

\begin{figure*}[t]
    \centering
    \includegraphics[clip, width=0.95\linewidth]{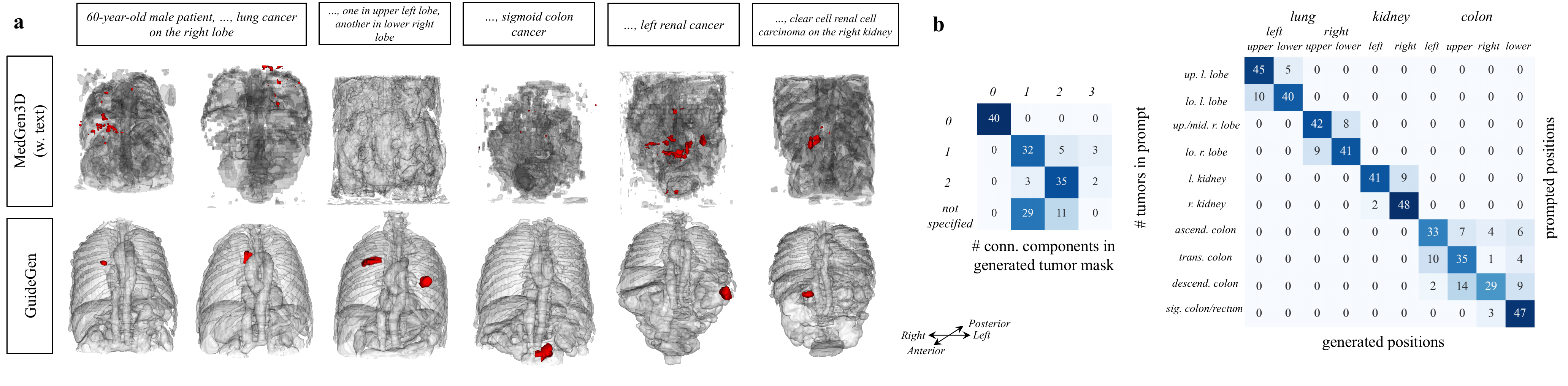}
    \caption{(a) Qualitative results of generated full-torso anatomical masks, with tumor masks masked in red. (b) Quantitative results evaluating GuideGen's mask-prompt alignment from two dimensions including the number of tumor and tumor location.}
    \label{fig:mask-prompt consistency}
\end{figure*}
\begin{table}[t] 
\Huge
    \centering
    \resizebox{0.7\linewidth}{!}{
    \begin{tabular}{l|c|cc|cc}
    \toprule
        \multirow{2}{*}{Mask Synthesizers} & Parameter& \multicolumn{2}{c|}{Full Anatomy}&\multicolumn{2}{c}{Tumor Only}  \\
        &Count(M)&LPIPS$\downarrow$&FID$\downarrow$&LPIPS$\downarrow$&FID$\downarrow$\\\midrule
        MedGen3D(w. text)&48.8&0.70&201&\textbf{0.29}&\underline{33.5}\\
        LDM&115.2& \underline{0.67} & \underline{98.6}&{0.30}&69.1\\
       Hu's&-&-&-&0.32&64.8\\
        GuideGen&51.5&\textbf{0.33} & \textbf{7.1}&\underline{0.29}&\textbf{27.9}\\
        \bottomrule
    \end{tabular}
    }
    \caption{Comparison of different semantic generation methods of full-torso anatomies and tumor-only binary masks. $\uparrow(\downarrow)$ indicates higher(lower) values are better. We use \textbf{bold} and \underline{underlined} metrics to indicate best and second bests.}
    \label{tab:mask}
\end{table}
\begin{table}[t]
\Huge
\setlength{\tabcolsep}{5mm}
    \centering
     \resizebox{\linewidth}{!}{
    \begin{tabular}{l|cc|c|ccc@{}}
    \toprule
        \multirow{2}{*}{Methods} & \multicolumn{2}{c|}{inference cond.} & Parameter& \multicolumn{3}{c}{CT generative metrics}\\
        & mask & text & Count(M)& LPIPS$\downarrow$ & FID$\downarrow$ & FVD$\downarrow$ \\\midrule
        Pinaya's & N&Y &196.1&0.465&92.0&2091\\
        GenerateCT & N & Y &288.7&0.521&151.7&4634\\ 
        MedSyn(text-only) &N & Y &115.1&0.396&50.0&2012\\\midrule
        MAISI &Y & N &166.5&0.406$_{\rm G}$, 0.393$_{\rm R}$&57.7$_{\rm G}$, 54.6$_{\rm R}$&1890$_{\rm G}$, 1791$_{\rm R}$\\
        Zhuang's &Y & N &81.2&0.327$_{\rm G}$, 0.335$_{\rm R}$&\underline{23.2}$_{\rm G}$, 39.1$_{\rm R}$&\underline{1094}$_{\rm G}$, 1366$_{\rm R}$\\
        MedDDPM & Y & N &80.5&0.354$_{\rm G}$, 0.342$_{\rm R}$ & 35.6$_{\rm G}$, 45.8$_{\rm R}$ & 1299$_{\rm G}$, 1619$_{\rm R}$\\\midrule
        MedSyn & Y & Y &115.1&\underline{0.304}$_{\rm G}$, \underline{0.282}$_{\rm R}$&28.1$_{\rm G}$, \underline{26.7}$_{\rm R}$&1450$_{\rm G}$, \underline{1288}$_{\rm R}$ \\ \midrule
        GuideGen & N & Y  &164.1&\textbf{0.248}$_{\rm G}$, \textbf{0.256}$_{\rm R}$ & \textbf{20.2}$_{\rm G}$, \textbf{19.4}$_{\rm R}$ & \textbf{791}$_{\rm G}$, \textbf{745}$_{\rm R}$\\
        \bottomrule
    \end{tabular}
    }
    \caption{Comparison studies of different image generation frameworks. Metrics suffixed with `G' or `R' separately denote input semantic guidance from GuideGen-generated masks and real masks. Note that GuideGen only rely on external textual inputs at inference.}
    \label{table:image}
\end{table}

\noindent\textbf{Generation quality} (1) \textbf{Mask quality}: To validate that GuideGen's ambiguity-reducing modeling is indeed superior to current methods based on continuous formulation, we choose MedGen3D~\cite{han2023medgen3d}, which is based on DDPM~\cite{ho2020denoising}, and LDM~\cite{rombach2021highresolution} as baselines and infuse them with textual information with cross-attention modules. Their generation results are rounded to the nearest integer to represent semantic information. We also compare with Hu's deformation method for tumor generation~\cite{hu2023label}. The results are shown in Tab.~\ref{tab:mask}. It is evident that GuideGen performs better than baselines by a large margin on anatomy generation. This is not surprising as the ambiguity problem mentioned in Sec.~\ref{sec:TCSS} significantly hinders the performance of baselines when the number of semantic classes $N$ is large. This problem becomes less prominent as $N$ drops, as shown by the last two columns, where GuideGen maintains comparable performance to baseline methods.
(2) \textbf{Image quality}: We compare GuideGen's image generation quality with a series of generative methods, including Pinaya's and MedDDPM~\cite{pinaya2022brain,medicalDDPM} for brain MRI generation, GenerateCT~\cite{hamamci2025generatect} and MedSyn~\cite{xu2024medsyn} for lung CT generation, Zhuang's~\cite{zhuang2023semantic} for abdominal CT generation and the generic medical generative framework MAISI~\cite{guo2024maisi}. We make minor modifications to the baseline methods, retrain them on the same datasets as GuideGen and fix the output CT volume size of all methods at $128^3$ to fit within our VRAM constraints. At inference, for methods that require a mask input, we use both GuideGen-generated and real masks as conditions to ensure a fair comparison in Tab.~\ref{table:image}. We can observe that GuideGen consistently outperforms existing methods in image quality across all metrics. Particularly, we notice that methods based solely on texts usually perform worse than those with semantic masks, as the highly diverse CT textures and intensity ranges among organs are difficult to recover without voxel-wise guidance, reflecting the necessity of our TCSS. Moreover, we see little discrepancy between metrics computed from CTs given by GuideGen-generated and real-world masks, reflecting that our TCSS can provide comparable quality to real masks. Quality results in Fig.~\ref{fig:results} also shows that GuideGen gives the most realistic and consistent CT generations.

\begin{table}[t]
\Huge
    \centering
     \resizebox{\linewidth}{!}{
    \begin{tabular}{l|ccccccccccc}
    \toprule
        \multirow{2}{*}{Methods} & \multicolumn{11}{c}{DSC$\uparrow$}\\
        & Spl. & Kid. & Liver & Sto. & Pan. & Lung & S.B. & Duo. & Colon & Heart & Avg.  \\\midrule
        MAISI &\underline{0.73} &\underline{0.72} &\underline{0.80} &\underline{0.60} &\underline{0.43} &\underline{0.84} &\underline{0.49} &0.35&0.55&\underline{0.40}&\underline{0.59} \\
        Zhuang's &0.43&0.43&0.62&0.36&0.21&0.69&0.37&0.23&0.26&0.24&0.38 \\
        MedDDPM &0.35&0.36&0.39&0.54&0.29&0.34&0.47&\textbf{0.43}&\underline{0.67}&0.22&0.41\\
        MedSyn&0.52&0.51&0.51&0.40&0.07&0.59&0.12&0.01&0.54&0.22&0.35 \\
        GuideGen&\textbf{0.75}&\textbf{0.72}&\textbf{0.90}&\textbf{0.63}&\textbf{0.46}&\textbf{0.84}&\textbf{0.51}&\underline{0.41}&\textbf{0.70}&\textbf{0.53}&\textbf{0.65} \\
        \bottomrule
    \end{tabular}
    }
    \caption{Image-mask alignment between generated CT and their full-torso anatomical guidance. We report the DSC scores on ten major thoracic and abdominal organs.}
    \label{tab:image-mask consistency}
\end{table}
\begin{table}[t]
\Huge
\setlength{\tabcolsep}{5mm}
    \centering
     \resizebox{0.75\linewidth}{!}{
    \begin{tabular}{l|ccccc}
    \toprule
        \multirow{2}{*}{Methods}  & \multicolumn{5}{c}{Accuracy$\uparrow$} \\
        & Age & Gender & Race & Tumor Loc. & Avg.  \\\midrule
        Pinaya's &0.06&0.35&0.10&0.17&0.17\\
        GenerateCT &0.07&0.21&0.44&0.03&0.19\\
        MedSyn&\underline{0.17}&\underline{0.74}&\underline{0.51}&\underline{0.47}&\underline{0.47} \\
        GuideGen &\textbf{0.39}&\textbf{0.90}&\textbf{0.60}&\textbf{0.89}&\textbf{0.69} \\
        \bottomrule
    \end{tabular}
    }
    \caption{Image-prompt alignment between generated CT and textual inputs. We report the accuracy of 4 dedicated classifiers on demographic and clinical features derived from input medical prompts.}
    \label{tab:image-text consistency}
\end{table}
\begin{table*}[t]
\Huge
    \centering
    \setlength{\tabcolsep}{6mm}
    \resizebox{\linewidth}{!}{
    \begin{tabular}{ll|c|cccccccccccccc}
    \toprule
        \multirow{2}{*}{Dataset}&\multirow{2}{*}{Method} &{No. Train} & \multicolumn{14}{c}{DSC$\uparrow$}\\
        &&{Cases}&Spleen & Kidneys & Liver & Sto. & Pan. & Adr. &Eso. & Aorta & IVC & Gall. & Duo.& Blad. &PV\&SV& Avg. \\\midrule
        \multirow{6}{*}{BTCV}&Real&24&0.92&0.79&0.94&0.86&0.7&0.6&0.71&0.89&0.81&0.52&-&-&0.52&0.74 \\\cmidrule[.5pt]{2-17}
        &MAISI &200&\underline{0.91} &0.89 &0.94 &0.80 &0.61 &0.44 &0.60 &0.84 &0.78 &0.29 &-&-&0.48 &0.69\\
        &Zhuang's &200&0.90 &0.88 &0.95 &0.83 &0.65 &0.54 &0.69 &\underline{0.88} &0.82 &\underline{0.49} &-&-&\underline{0.56}&0.74  \\
        &MedDDPM &200&0.92 &0.90 &0.95 &\underline{0.87} &\underline{0.66} &0.54 &0.68 &0.88 &0.84 &0.39 &-&-&0.49 &\underline{0.74}  \\
        &MedSyn &200& 0.89 &\underline{0.90} &\underline{0.96} &0.81 &0.65 &\underline{0.56} &\underline{0.70} &0.86 &\underline{0.86} &0.39 &-&-&0.32 &0.72 \\
        &GuideGen &200& \textbf{0.96} &\textbf{0.91} &\textbf{0.98} &\textbf{0.90} &\textbf{0.76} &\textbf{0.62} &\textbf{0.74} &\textbf{0.92} &\textbf{0.90} &\textbf{0.49} &-&-&\textbf{0.57} &\textbf{0.79} \\
        \midrule
        \multirow{6}{*}{AMOS}&Real&240&0.95&0.94&0.96&0.89&0.81&0.67&0.78&0.92&0.87&0.72&0.76&0.82&-&0.84\\\cmidrule[.5pt]{2-17}
        &MAISI &200&0.83 &0.84 &0.91 &\underline{0.74} &0.60 &0.50 &0.60 &0.81 &0.71 &0.34 &0.53 &0.43 &-&0.65\\
        &Zhuang's &200&0.85 &0.87 &0.90 &0.66 &0.63 &0.46 &0.61 &0.82 &0.73 &0.26 &\underline{0.55} &0.62 &-&0.66  \\
        &MedDDPM &200& \underline{0.86} &0.89 &0.90 &0.74 &0.61 &0.46 &0.61 &\underline{0.86} &\underline{0.76} &\underline{0.43} &0.55 &0.60 &-&\underline{0.69} \\
         &MedSyn &200& 0.85 &\underline{0.90} &\underline{0.91} &0.70 &\underline{0.64} &\underline{0.52} &\underline{0.62} &0.80 &0.69 &0.21 &0.53 &\underline{0.64} &-&0.67 \\
        &GuideGen&200&\textbf{0.95} &\textbf{0.92} &\textbf{0.95} &\textbf{0.90} &\textbf{0.70} &\textbf{0.52} &\textbf{0.73} &\textbf{0.88} &\textbf{0.82} &\textbf{0.60} &\textbf{0.67} &\textbf{0.72} &-&\textbf{0.78} \\
        \bottomrule
    \end{tabular}
    }
    \caption{Segmentation performance on multi-organ segmentation tasks using real or synthetic data from each framework.}
    \label{table:mos}
\end{table*}

\noindent\textbf{Conditional alignment} 
(1) \textbf{Image-mask alignment}: Image-mask alignment can be quantitatively measured by comparing mask predictions from pretrained segmentation models~\cite{wasserthal2023totalsegmentator} on generated images with the actual masks given, with a higher alignment being a higher segmentation accuracy. We report the DSC of 10 major organs between predicted masks and input semantic condition in Tab.~\ref{tab:image-mask consistency}. It can be seen that GuideGen behaves desirably for semantic correspondence, achieving a 6\% gain on DSC compared to its nearest competitor.
(2) \textbf{Image-prompt alignment}: To quantitatively evaluate the alignment between generated images and text prompt conditions, similar to~\cite{gu2023biomedjourney,hamamci2025generatect}, we use a series of pretrained classifiers on certain features to judge whether the generated images faithfully exhibit features specified in medical prompts. As shown in Tab.~\ref{tab:image-text consistency}, images generated by our framework faithfully reflect the gender information and tumor location while outperforming other text-conditioned frameworks in preserving other textual features.
(3) \textbf{Mask-prompt alignment}: We measure mask-prompt alignment qualitatively using mask synthesizers in MedGen3D (adapted by adding textual constraints to its generation backbone) and GuideGen in Fig.~\ref{fig:mask-prompt consistency}(a) by varying the number and location of tumor sites in our text prompt. We also quantify GuideGen's mask-prompt alignment in Fig.~\ref{fig:mask-prompt consistency}(b). It can be seen that our categorical modeling in TCSS not only generates masks of accurate shape and structure, but also preserves the spatial relationships between tumor and organs destined in the input prompts.

\noindent\textbf{Downstream Usability}
To assess the usability in synthesizing datasets for segmentation, we use nnU-Net~\cite{isensee2021nnu} to train segmentation models with samples generated by different mask-based generative frameworks. Sample synthesis is based on the masks and/or textual prompts in the validation split of our TCIA and RJ datasets.
(1) \textbf{Multi-organ segmentation}: As shown in Tab.~\ref{table:mos}, GuideGen's synthetic samples yield segmentation performance comparable to real data and superior to samples from baseline frameworks, highlighting our method's generation quality.
(2) \textbf{Tumor segmentation}: For tumor segmentation (Tab.~\ref{table:tumor}), GuideGen-generated lung tumors yield results competitive with real data, while on other tumor segmentation tasks, GuideGen's generated samples prove more beneficial than those from other baselines.

\subsection{Ablation studies}
\begin{table}[t]
\Huge
    \centering
    \setlength{\tabcolsep}{5mm}
    \resizebox{1\linewidth}{!}{
    \begin{tabular}{l|c|cccc|cccc}
    \toprule
        \multirow{2}{*}{Method} &{No. Train} & \multicolumn{4}{c|}{DSC$\uparrow$} & \multicolumn{4}{c}{$\text{HD}_{95}\downarrow$}\\
         &Cases& LU & CO  & KI & Avg.  & LU & CO  & KI & Avg.\\\midrule
        Real&50/100/313 &0.69 &0.47 &0.72 &0.63 &11.9 &212.7 &70.9 &98.5 \\\midrule
        MAISI&200 &\underline{0.48} &0.10 &0.24 &0.27 &\underline{31.7} &284.9 &293.5 &203.4   \\
        Zhuang's&200 & 0.12 &0.07 &0.13 &0.11 &716.0 &274.1 &511.5 &500.5   \\
        MedDDPM&200 &0.10 &0.09 &0.29 &0.16 &776.0 &273.6 &\underline{119.4} &389.7  \\
        MedSyn&200 &0.44 &\underline{0.11} &\underline{0.39} &\underline{0.31} &33.1 &\underline{267.4} &309.0 &\underline{203.2}  \\
        GuideGen &200&\textbf{0.71} &\textbf{0.21} &\textbf{0.64} &\textbf{0.52} &\textbf{8.4} &\textbf{227.0} &\textbf{84.5} &\textbf{106.6} \\
        \bottomrule
    \end{tabular}
    }
    \caption{Segmentation performance on 3 tumor segmentation tasks using real or synthetic cases from each method.}
    \label{table:tumor}
\end{table}
\begin{table}[t] 
    \centering
    \Huge
    \resizebox{0.9\linewidth}{!}{
    \begin{tabular}{l|cc|cc}
    \toprule
        {\centering Experiment Setup}&LPIPS$\downarrow$&FID$\downarrow$&Avg. DSC$\uparrow$&Avg. Acc.$\uparrow$\\\midrule
        \textit{w.o. mask input}&0.42&54.3&-&0.32\\
        \textit{LDM-generated mask input}&0.34&38.4&0.34&0.40\\
        \textit{MedGen3D-generated mask input}&0.36&39.0&0.23&0.41\\
        \midrule
        \textit{w.o. knowledge injection}&0.26&21.7&0.25&0.57\\
        \midrule
        \textit{w.o. anatomy preservation}&0.27&32.4&0.40&0.61\\
        \textit{w.o. HDR accommodation}&0.33&40.9&0.36&0.64\\\midrule
        GuideGen&\textbf{0.25}&\textbf{20.2}&\textbf{0.52}&\textbf{0.69}\\
        \bottomrule
    \end{tabular}
    }
    \caption{Ablations studies on GuideGen. DSC and Acc. are separately evaluated and averaged on 3 downstream tumor segmentation tasks and 4 image-prompt alignment tasks.}
    \label{tab:ablations}
\end{table}
\begin{figure}[t]
    \centering
    \includegraphics[width=1\linewidth,]{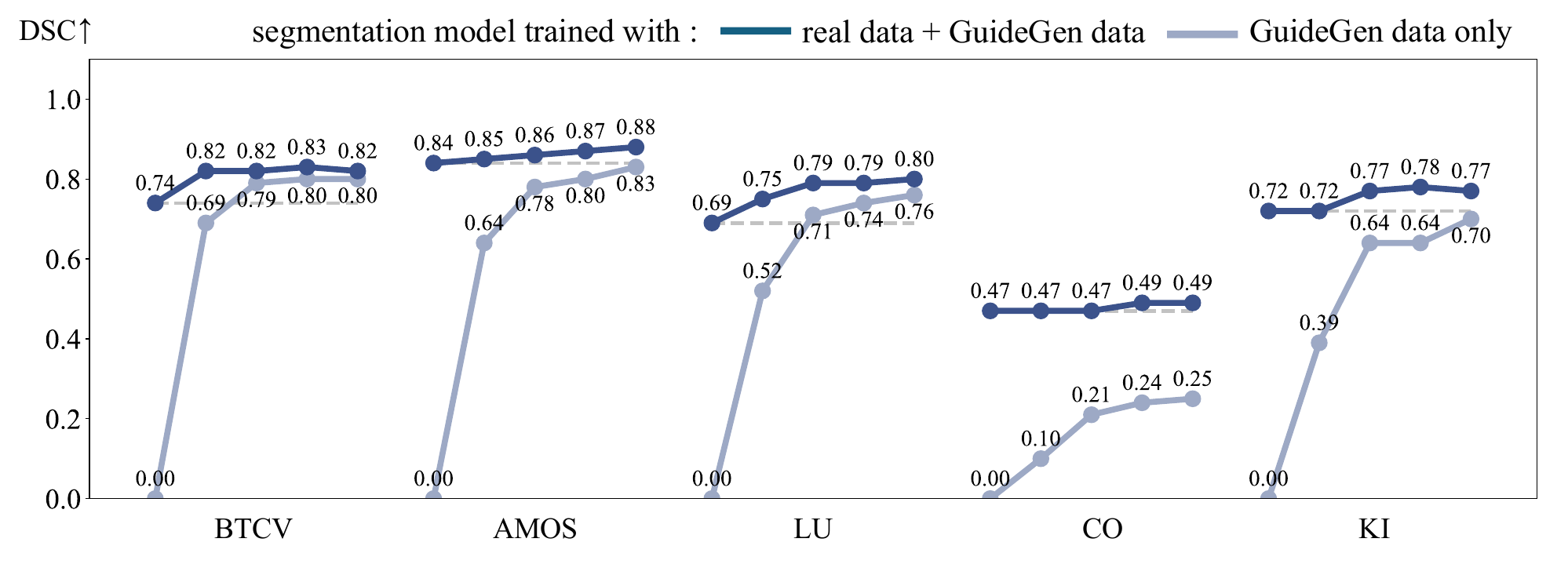}
    \caption{Segmentation performance using different number (0, 100, 200, 500, 1K) of GuideGen-generated samples as augmentation. Darker and lighter solid lines separately denote segmentation model trained with or without real data.}
    \label{fig:aug}
\end{figure}
(1) \textbf{Mask Synthesizers}: By replacing the mask synthesizer (Stage-I) of our GuideGen with other off-the-shelf mask generators, we can ablate on the CT generative performance using different mask synthesizers, as illustrated by the first 3 rows and the last row of Tab.~\ref{tab:ablations}. It can be seen that CTs generated with semantic guidance from TCSS performs best, and a lack of semantic guidance degrades performance drastically. Comparing Tab.~\ref{tab:ablations} with Tab.~\ref{tab:mask}, we can see an interesting trend that generated CT quality is roughly in line with the quality of mask inputs. (2) \textbf{Knowledge Injection}: The superiority of knowledge injection as opposed to conventional cross-attention guidance is illustrated in rows 4, 7 of Tab.~\ref{tab:ablations}, with a significant boost of alignment and downstream segmentation metrics indicating that knowledge injection better preserves textual features. (3) \textbf{Anatomy-aware HDR Autoencoder}: The last 3 rows in Tab.~\ref{tab:ablations} also shows a clear boost in generative metrics by integrating HDR accommodation and anatomy preservation to CT autoencoders for full-torso generation, which also promotes downstream usability as the image quality becomes more desirable. (4) \textbf{Generation quantity}: From Fig.~\ref{fig:aug}, we can clearly see a trend that with a larger quantity of generated samples, downstream segmentation performance can be better boosted across all tasks of choice. Also, for tasks with a simpler background (like LU), the performance of segmentation models trained using only GuideGen-generated samples can be comparable to or even better than those trained with real cases only.

\section{Conclusion}
In this paper, we have demonstrated a novel framework for generating paired 3D anatomies and CT images covering the entire torso, conditioned on a structured medical prompt. Specifically, our GuideGen outperforms a range of existing methods under comprehensive evaluation, including generative quality, alignment with textual and semantic guidance, and usability in training semantic segmentation models. Overall, our approach not only advances the state of the art in text-guided 3D CT generation but also lays a foundation for more robust training dataset synthesis and augmentation for medical image analysis.
\textbf{Limitations} While GuideGen achieves strong performance to synthesize full-torso semantics and CTs from structured prompts given by Large Language Models, its cannot directly generate plausible images from free-text inputs. We consider this as a potential future direction for enabling more versatile control.

\section{Acknowledgements}
This work was supported by National Natural Science Foundation of China (NSFC 62301311).

\small
\bibliography{aaai2026}

\clearpage
\appendix

\section{Datasets and Implementation Details}
\subsection{Dataset Details}
\label{sec:appendix data}

\begin{figure*}[h]
    \centering
    \includegraphics[width=\linewidth]{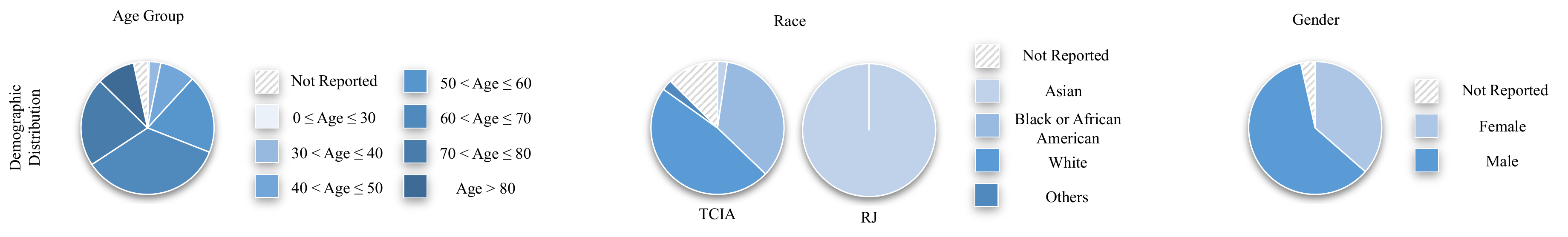}
    \caption{\textbf{Demographics distribution for TCIA and RJ datasets} used for generative training as well as quality and alignment evaluations.}
    \label{fig:demographics}
\end{figure*}

\paragraph*{Training datasets} We compiled 12 TCIA~\cite{clark2013cancer} datasets on three common types of cancer (lung, kidney and colon). To supplement the scarcity of publicly available colon cancer multimodal (paired CT and clinical information) dataset, we also include one private real-world dataset, dubbed RJ, for colorectal cancer. The preprocessing of all the collected unstandardized samples include the following four steps: (a) \textbf{Data cleansing}: All of the collected CT samples are reoriented into RAS orientation and resampled to isotopic spacing, with duplicate, impaired, partial DICOM series identified and removed. (b) \textbf{Label Extraction}: CT images in the collected dataset are segmented by TotalSegmentator~\cite{wasserthal2023totalsegmentator} and pretrained models~\cite{isensee2021nnu} on their respective tumor types to retrieve both multi-organ and tumor segmentation masks; their corresponding clinical information are retrieved using a private medical GPT in a certain format: \textbf{\textit{The patient is \{demographics group\}. In this imaging, the patient's condition is described as \{clinical information\}}}. (c) \textbf{Noise Removal}: Organ labels belonging to the same anatomical class (\textit{e.g.} individual rib label and the anatomical class ribs) are merged into one semantic label to mitigate noise in the multi-organ labels. Samples with no or erroneous (tumor predictions outside organ prediction) tumor mask predictions are discarded. (d) \textbf{Dataset Preparation}: We randomly split the processed images into 4534 training cases and 1179 validation cases. Specifically, we use the CT, clinical information and masks in the training split to train GuideGen as well as baseline models, detailed description of each training dataset of choice are listed in Tab.~\ref{tab:dataset}, and we showcase the distribution of each demographics group in TCIA and RJ datasets in Fig.~\ref{fig:demographics}.
\begin{table*}[h]
    \centering
    \resizebox{\linewidth}{!}{
    \begin{tabular}{l|cccc}
    \toprule
    Dataset Name&Description&No. Patients&No. Cases&No. Cases after pp.\\\midrule
    CMB-CRC~\cite{Cancer_Moonshot_Biobank2022}&Colorectal Cancer&54&237&50\\
    TCGA-READ~\cite{Kirk2016}&Rectum Cancer	&3&34&4\\
    TCGA-COAD~\cite{Kirk2016-ty}&Colon Cancer	&25&93&21\\
    RJ&Colorectal Cancer&4005&4418&3235\\\midrule
    CPTAC-CCRCC~\cite{ccrcc-zv}&Renal Cancer&262&727&248\\
    TCGA-KICH~\cite{Linehan2016-qd}&Renal Cancer&15&109&26\\
    TCGA-KIRC~\cite{Akin2016-uc}&Renal Cancer&267&2654&734\\
    TCGA-KIRP~\cite{Linehan2016-uj}&Renal Cancer&33&376&56\\\midrule
    CPTAC-LSCC~\cite{cptac-hu}&Pulmonary Cancer&212&238&236\\
    CPTAC-LUAD~\cite{cptac-fz}&Pulmonary Cancer&244&242&71\\
    CMB-LCA~\cite{Cancer_Moonshot_Biobank2022-ra}&Pulmonary Cancer&97&696&64\\
    TCGA-LUAD~\cite{Albertina2016-nm}&Pulmonary Cancer&69&624&120\\
    TCGA-LUSC~\cite{Kirk2016-wa}&Pulmonary Cancer&37&279&92\\
    \bottomrule
    \end{tabular}
    }
    \caption{\textbf{Datasets used for generative training, quality evaluation and alignment measurement}. ``RJ" is our private dataset on colorectal cancer while ``No. Cases after pp." is short for ``The number of cases after preprocessing".}
    \label{tab:dataset}
\end{table*}

\paragraph*{Inference datasets} We we choose the abdominal split of Multi-Atlas Labeling Beyond the Cranial Vault (BTCV)~\cite{landman2015miccai} and Abdominal~\cite{ji2022amos} Multi-Organ Segmentation (AMOS) datasets and evaluate the segmentation performance on some of their organ labels that overlap with our interest. Additionally, for tumor segmentation, we select two Medical Segmentation Decathlon (MSD)~\cite{antonelli2022medical} datasets along with the Kidney Tumor Segmentation (KiTS)~\cite{heller2023kits21} dataset, separately for pulmonary, colorectal and renal cancer. Their details are documented in Tab.~\ref{tab:dataset2}. We also downsamples the training cases in each inference dataset to $128^3$ to ensure a fair comparison with the $128^3$ volumes generated by GuideGen and other baselines.
\begin{table*}[t]
    \centering
    \resizebox{\linewidth}{!}{
    \begin{tabular}{l|cccc}
    \toprule
    Dataset Name&Description&No. Train cases& No. Inference cases& Semantic Classes Reported\\\midrule
    BTCV~\cite{landman2015miccai}&Abdominal Organs&24&6&Spl., Kid., Liver, Sto., Pan., Adr., Eso., Aorta, IVC, Gall., PV\&SV\\
    AMOS~\cite{ji2022amos}&Abdominal Organs&240&60&Spl., Kid., Liver, Sto., Pan., Adr., Eso., Aorta, IVC, Gall., Duo., Blad.\\\midrule
    MSD-Lung~\cite{antonelli2022medical}&Pulmonary Cancer&40&10&Lung Tumor\\
    MSD-Colon~\cite{antonelli2022medical}&Colon Cancer&80&20&Colon Tumor\\
    KiTS~\cite{heller2023kits21}&Renal Cancer&210&90&Kidney Tumor\\
    \bottomrule
    \end{tabular}
    }
    \caption{\textbf{Datasets used for downstream segmentation evaluation}. Abbreviations: Spl.: Spleen, Kid.: Kidneys, Sto.: Stomach, Pan.: Pancreas, Adr.: Adrenal Glands, Eso.: Esophagus, IVC: Inferior Vena Cava, Gall.: Gallbladder, PV\&SV: Portal Vein and Splenic Vein, Duo.: Duodenum, Blad.: Urinary Bladder.}
    \label{tab:dataset2}
\end{table*}
\subsection{Implementation Details} 
\label{sec:appendix imp}
\paragraph*{GuideGen} In Stage-I, we use TCSS to generate 19 major pleural, abdominal and pelvic structure\footnote{The 19 major anatomical structure of choice are: Lung, Heart, Ribs, Vertebra, Esophagus, Aorta, Inferior Vena Cava, Liver, Spleen, Stomach, Duodenum, Gallbladder, Splenic Vein and Portal Vein, Kidneys, Adrenal Glands, Pancreas, Small Bowel, Colon (with rectum) and Urinary Bladder, covering major anatomies throughout the torso.} and 1 tumor semantics. The reason for not choosing more organ labels lies largely in that other anatomies suffer from relatively poor multi-organ segmentation accuracies. It should be noted that our framework should be simulate as many anatomies as possible if the quality of training data is ensured. The model channels in each layer of our backbone U-Net network are 32, 64, 128, 256, respectively, with 16$\times$ downsampling at the U-Net bottleneck. Each layer consists of two ResNet convolution blocks and we add a cross-attention layer to the first and third layer to integrate textual information. Additionally, the knowledge injection module utilizes a mask/image query of length 64 and latent dimension 8 to extract task-specific information.

In Stage-II, our autoencoder operates on a latent dimension of $64^3$, or compress the input $256^3$ CT image by 4$\times$. The frame discriminator $\mathbf{D}_f$ in the loss computation chooses 4 slices for discriminative loss computation in each iteration. We also set a min-max normalization for CT images prior to their input into the autoencoder model: CT images are linearly rescaled in such a way that the CT intensities of the range [-1500, 1500]HU resides in the interval [-1, 1] after rescaling. In this way, we can recover CT intensities from the decoder half of our autoencoder after diffusion generation in Stage-III by multiplying the result by the multiplier 1500.

In Stage-III, we first retrieve the semantic information for organ and tumor masks individually: Firstly, for 19 organ labels $l_A=\{l_A^1,\cdots,l_A^{19}\}$, 1 tumor label $l_T$ and 1 background label $l_B$, the final step of TCSS sampling trajectory will output a probability vector $\hat{\mathbf{x}}_0\in[0,1]^{21\times H\times W\times D}$. $H$, $W$ and $D$ are spatial dimensions, which all equals 128 in our setting. Secondly, we perform \verb|argmax| operation on the channel dimension to retrieve these semantic masks. Specifically, we find that it is better to retrieve the semantic labels via the following way separately for organ $\mathbf{m}_A$ and tumor semantics $\mathbf{m}_T$: 
\begin{equation}
    \mathbf{m}_A=\mathop{argmax}_{l_A,l_B}\hat{\mathbf{x}}_0;\;
    \mathbf{m}_T=\delta(\mathop{argmax}_{l_A,l_T,l_B}\hat{\mathbf{x}}_0, l_T).
\end{equation}
That is, we discard the probabilities of tumor label when retrieving organ semantics. This is done so that the diffusion model can accurately pin each tumor type on a specific organ and to raise special awareness to our latent-guided generator by separating tumor mask as a separate semantic channel. $\delta(x,y)$ here is the impulse function that takes 1 on the entry $i$ when $x[i]$ equals $y$ and 0 otherwise. These semantic latents ($\mathbf{m}_A$ and $\mathbf{m}_T$) are then resampled to a desired output resolution (in our case it is 64 following the latent spatial resolution of our anatomy-aware HDR autoencoder), along with the image feature encoded by the encoder half of our autoencoder are then concatenated to a latent vector with a feature dimension of 16, which is fed into a backbone U-Net with the same structure and knowledge injection module as that described in TCSS. Loss is computed between the U-Net output and the noised image latent.

\paragraph*{Baseline models} We follow the official implementations available online for each baseline method, and adjust their model parameters (mainly the channel numbers for each convolutional layer) so that they can generate $128^3$ volumes under our physical setting (NVIDIA RTX 4090 GPUs). We did not use baseline models to generate $256^3$ volumes here because it would compromise some models' performance drastically and instead we adapt GuideGen to generate $128^3$ volumes to compare with them fairly (see the Experiment section in the main text). We also skip the final super-resolution stage in some baselines~\cite{xu2024medsyn,hamamci2025generatect} since it has no relevance to conditional generation and can be added to any generative framework as a postprocessing module.

\section{Supplementary Results}
\label{sec:suppl results}
This section provides tables and figures that serve as a supplement to the main results displayed in the main article. 
\subsection{Qualitative Results}
\begin{figure*}
    \centering
    \includegraphics[width=\linewidth]{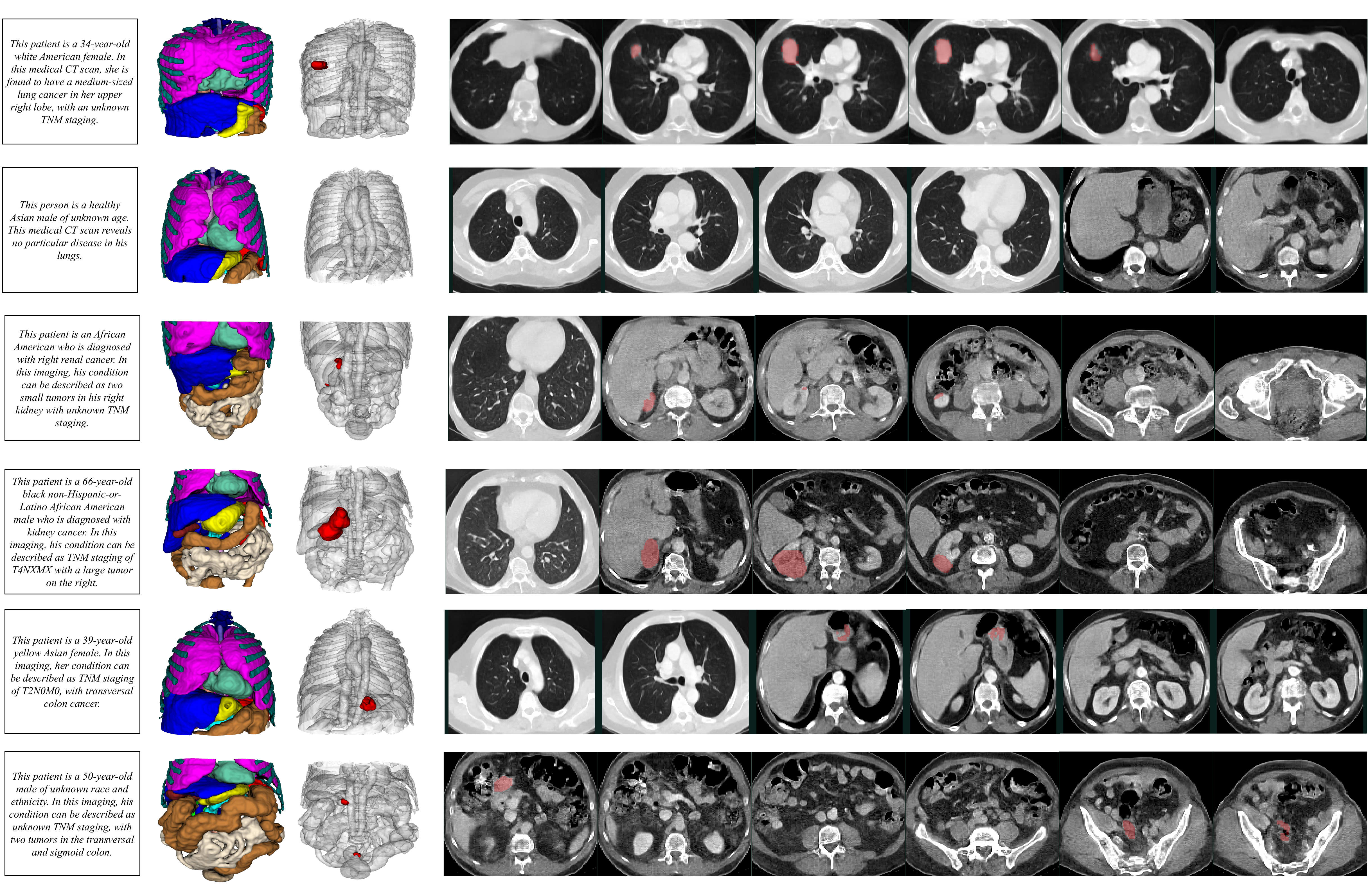}
    \caption{\textbf{Masks and Images generated by GuideGen for qualitative assessment}. 6 slices are shown in each case along with the multi-organ anatomies and tumor semantics (if any). We show the input medical prompts on the left and mark the tumor labels in the CT slices shown in red.}
    \label{fig:qualityshowcase}
\end{figure*}

\begin{figure*}[t]
    \centering
    \includegraphics[width=\linewidth]{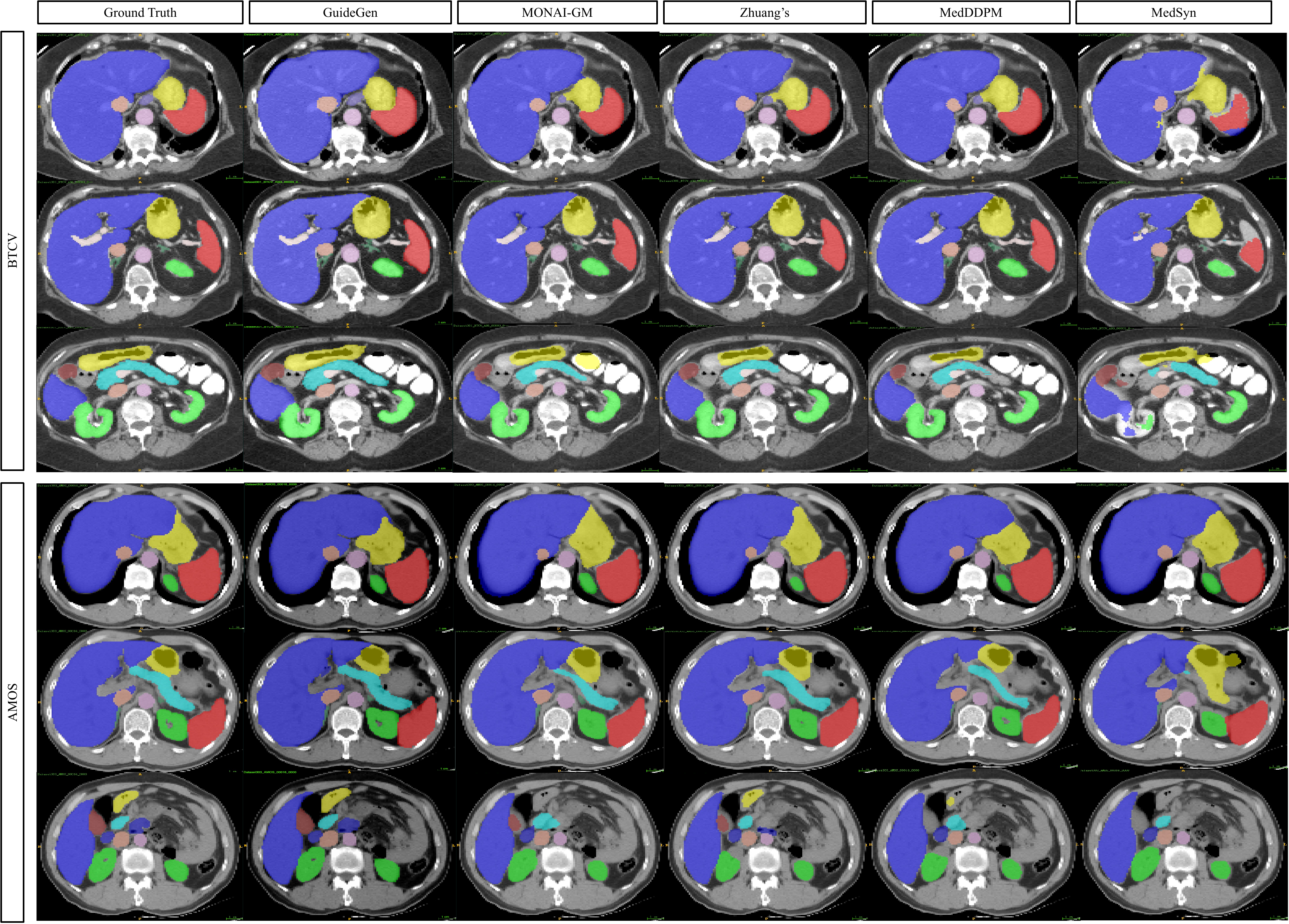}
    \caption{\textbf{Qualitative results for multi-organ segmentations on BTCV and AMOS datasets}. Three slices are shown in each case displayed.}
    \label{fig:mos}
\end{figure*}

\begin{figure*}[t]
    \centering
    \includegraphics[width=\linewidth]{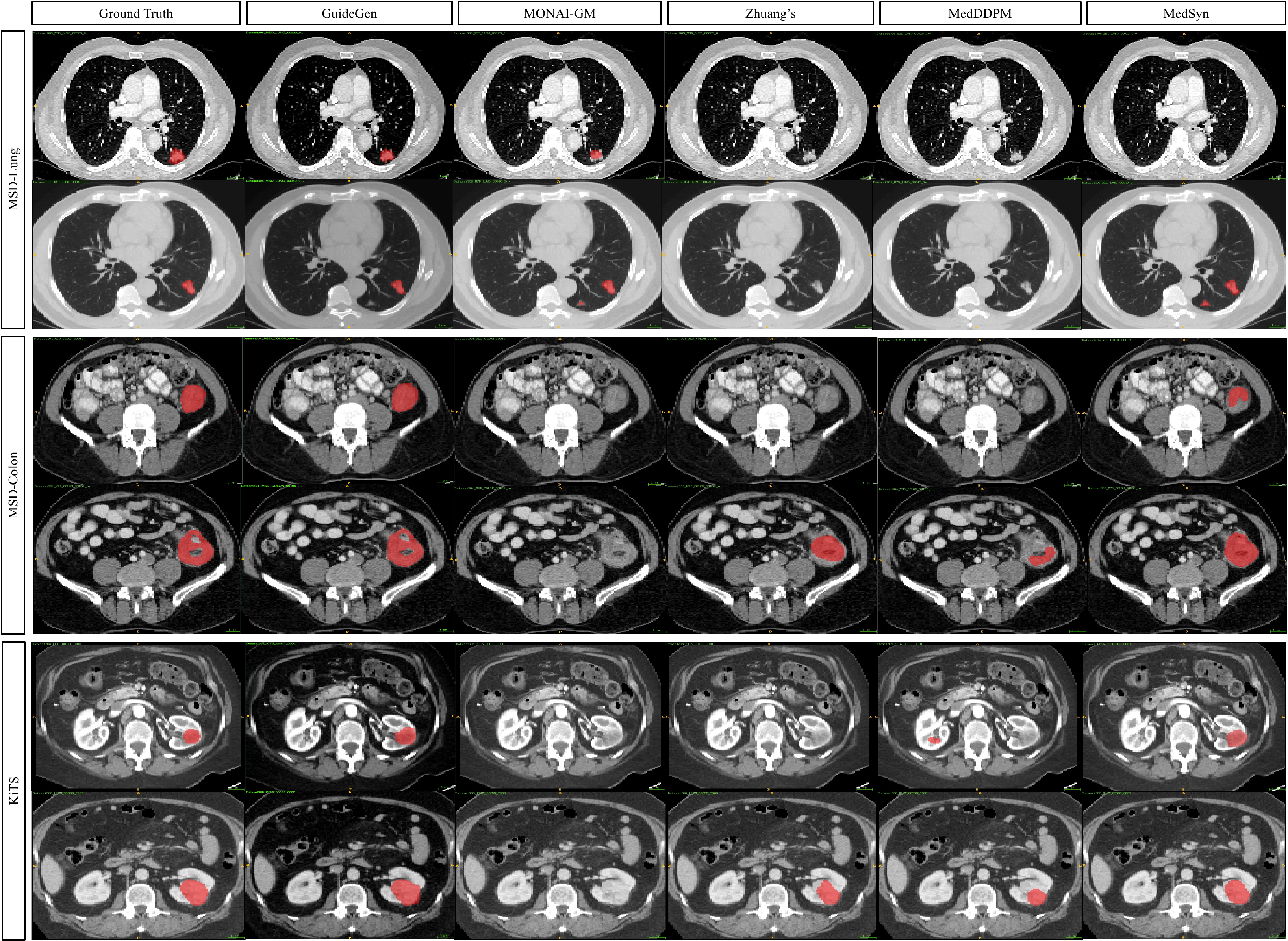}
    \caption{\textbf{Qualitative results for tumor segmentations on LU, CO and KI datasets}. One slice is chosen for each case, which is centered at the largest transversal section of the tumor.}
    \label{fig:ts}
\end{figure*}

In this section, we first showcase GuideGen's generative performance of both unhealthy and healthy samples using randomly chosen demographic and clinical descriptions from TCIA and RJ datasets in Fig.~\ref{fig:qualityshowcase}. Secondly, we display multi-organ segmentation results and tumor segmentation results separately in Fig.~\ref{fig:mos} and Fig.~\ref{fig:ts}. For slices located in the pleural region, we display them using a normalization window at [-975, -225]HU, while for slices in the abdominal and pelvic regions, we display them using a normalization window at [-50, 150]HU as recommended in~\cite{seeram2015computed}. We generate mask volumes $\hat{\textbf{m}}$ at a resolution of $128^3$ and use trilinear interpolation to resample them to $64^3$ for CT volume generation. The generated $256^3$ CT volume $\hat{\textbf{v}}$ is then paired with $\hat{\textbf{m}}$ upsampled to the same spatial resolution as $\hat{\textbf{v}}$ as generated pair.
\subsection{Quantitative Results}
We display standard variances of each metric used in multi-organ segmentation and tumor segmentation evaluations separately in Tab.~\ref{tab:std1} and Tab.~\ref{tab:std2} in this section. We also report the validation accuracy and AUROC of demographic or clinical classifiers used to assess image-prompt alignment in Tab.~\ref{tab:classifier}. They are implemented based on the encoder half of U-Net 3D networks~\cite{cciccek20163d} and trained on the training split of TCIA and RJ datasets. Classifier classes are chosen following the demographic classes in Fig.~\ref{fig:demographics} and tumor locations in Tab.~\ref{tab:dataset}. Metrics in Tab.~\ref{tab:classifier} are computed on the validation split of TCIA and RJ datasets. 

\subsection{Comparison with Tumor Inpainting Methods for Segmentation Augmentations}
Finally, it has come to our knowledge that in addition to full-image generation like GuideGen and baseline methods described in the main paper, there has been a growing interest in using image inpainting to synthesize only lesion or tumor sites, such as DiffTumor~\cite{chen2024towards}, FreeTumor~\cite{wu2024freetumor} and LesionDiffusion~\cite{tian2025lesiondiffusion}. Here, we hope to first emphasize that GuideGen, enabled by TCSS, can synthesize not only tumor masks, but also full anatomies and that it learns the tumor distribution through learning, rather than random deformation as per current practices. Nevertheless, to illustrate that GuideGen is also capable of synthesizing high-quality tumors, we perform segmentation pretraining using \textbf{1)} real CTs and inpainted tumor patches via various tumor generation pipelines and \textbf{2)} GuideGen-generated whole CT volumes and present the results in Tab.~\ref{tab:difftumor}. It is evident that GuideGen can synthesize faithful tumor sites that aid for downstream tasks and are comparable or even superior to two baseline tumor inpainting methods' synthesized cases, even if these inpainting methods comes with task-specific designs. This comparison further highlights GuideGen's competence as a segmentation dataset generator.

\section{Limitation and Future Directions}
Although GuideGen currently can achieve downstream segmentation dataset construction and augmentation, its generation pipeline contains three-stages, which can make adaptation to other modalities or out-of-distribution datasets time-consuming. An end-to-end model would certainly be more preferable, but due to the ambiguity problem mentioned in the Method section in the main text, using a single diffusion model with continuous modeling may be insufficient to fully grasp the exact discrete nature of labels. Future works may revolve around finding more efficient way to represent discrete labels, which may include using codebooks to encode adjacent labels into distant vectors in high-dimensional space, thereby circumventing the issue of ambiguity. Another limitation of our work is that the input prompt, despite having no constraint on the words of choice, needs to be input with a certain format to draw the model's attention. A possible future direction can be using vision-language foundation models~\cite{carvalhido2025stress} to align features extracted from structure-free inputs with encoded image latents.

\begin{table*}[ht]
\Huge
    \centering
    \setlength{\tabcolsep}{5mm}
    \resizebox{\linewidth}{!}{
    \begin{tabular}{ll|cccccccccccccc}
    \toprule
        \multirow{2}{*}{Dataset}&\multirow{2}{*}{Method} & \multicolumn{14}{c}{Std. DSC}\\
        &&Spl. & Kid. & Liver & Sto. & Pan. & Adr. &Eso. & Aorta & IVC & Gall. & Duo.& Blad. &PV\&SV& Avg. \\\midrule
        \multirow{6}{*}{BTCV}&Real&0.295 &0.332 &0.132 &0.224 &0.271 &0.191 &0.180 &0.318 &0.365 &0.326 &-&-&0.208 &0.258  \\\cmidrule[.5pt]{2-16}
        &MONAI-GM &0.013 &0.018 &0.011 &0.090 &0.098 &0.070 &0.124 &0.073 &0.077 &0.310 &-&-&0.116 &0.091 \\
        &Zhuang's &0.024 &0.017 &0.010 &0.046 &0.089 &0.049 &0.093 &0.027 &0.041 &0.394 &-&-&0.156 &0.086   \\
        &MedDDPM &0.014 &0.014 &0.010 &0.049 &0.107 &0.072 &0.102 &0.061 &0.044 &0.352 &-&-&0.141 &0.088   \\
        &MedSyn &0.015 &0.012 &0.007 &0.027 &0.113 &0.086 &0.111 &0.062 &0.031 &0.398 &-&-&0.173 &0.094 \\
        &GuideGen & 0.012 &0.004 &0.010 &0.019 &0.094 &0.055 &0.086 &0.033 &0.028 &0.440 &-&-&0.035 &0.074  \\
        \midrule
        \multirow{6}{*}{AMOS}&Real&0.065 &0.068 &0.044 &0.172 &0.198 &0.172 &0.168 &0.133 &0.163 &0.292 &0.172 &0.256 &-&0.146 \\\cmidrule[.5pt]{2-16}
        &MONAI-GM &0.170 &0.157 &0.166 &0.176 &0.209 &0.187 &0.166 &0.166 &0.170 &0.292 &0.187 &0.270 &-&0.176 \\
        &Zhuang's &0.127 &0.100 &0.082 &0.190 &0.224 &0.213 &0.192 &0.156 &0.175 &0.332 &0.213 &0.228 &-&0.167   \\
        &MedDDPM & 0.062 &0.043 &0.062 &0.199 &0.192 &0.204 &0.186 &0.090 &0.123 &0.304 &0.204 &0.231 &-&0.141  \\
         &MedSyn & 0.070 &0.045 &0.087 &0.103 &0.172 &0.149 &0.163 &0.085 &0.131 &0.281 &0.149 &0.252 &-&0.127 \\
        &GuideGen&0.021 &0.056 &0.021 &0.099 &0.195 &0.130 &0.277 &0.060 &0.108 &0.199 &0.213 &0.297 &-&0.140  \\
        \bottomrule
    \end{tabular}}
    \caption{\textbf{Segmentation variance on multi-organ segmentation tasks using real or synthetic cases from each method}. The significance of our improvement over the second-best framework is tested using a two-sided $t$-test, with a $p$-value of 0.016 and 0.001 respectively for two datasets. Abbreviations: Spl.: Spleen, Kid.: Kidneys, Sto.: Stomach, Pan.: Pancreas, Adr.: Adrenal Glands, Eso.: Esophagus, IVC: Inferior Vena Cava, Gall.: Gallbladder, Duo.: Duodenum, Blad.: Urinary Bladder, PV\&SV: Portal Vein and Splenic Vein. Some of the semantic classes reported are not consistent with the original dataset since they are merged as described in~\ref{sec:appendix imp}.}
    \label{tab:std1}
\end{table*}

\begin{table}[ht]
    \centering
    \resizebox{1\linewidth}{!}{
    \begin{tabular}{l|cccc|cccc}
    \toprule
        \multirow{2}{*}{Method} & \multicolumn{4}{c|}{Std. DSC} & \multicolumn{4}{c}{Std. $\text{HD}_{95}$}\\
         & LU & CO  & KI & Avg.  & LU & CO  & KI & Avg.\\\midrule
        Real &0.22 &0.38 &0.32 &0.31 &20.9 &394.7 &242.0 &219.2  \\\midrule
        MONAI-GM &0.23 &0.14 &0.33 &0.23 &29.6 &403.1 &497.1 &309.9    \\
        Zhuang's & 0.12 &0.19 &0.27 &0.19 &28.8 &408.5 &483.6 &307.0    \\
        MedDDPM &0.21 &0.19 &0.38 &0.26 &21.2 &409.0 &275.9 &235.4   \\
        MedSyn &0.35 &0.18 &0.29 &0.27 &45.5 &411.4 &426.0 &294.3   \\
        GuideGen &0.16 &0.21 &0.45 &0.27 &16.6 &391.6 &312.7 &240.3  \\
        \bottomrule
    \end{tabular}}
    \caption{\textbf{Segmentation variance on tumor segmentation tasks of different organs using real or synthetic cases from each method}. The significance of our improvement over the second-best framework is tested using a two-sided $t$-test, with a $p$-value of 0.002, 0.01 and 0.06 respectively for three datasets. When there are no tumor predictions, HD$_{95}$ is set to be 1000 to avoid $\infty$ values for a valid comparison. Abbreviations: LU: pulmonary cancer, CO: colorectal cancer, KI: renal cancer.}
    \label{tab:std2}
\end{table}
\begin{table}[t]
    \centering
    \resizebox{1\linewidth}{!}{
    \begin{tabular}{l|cc|c}
    \toprule
        \multirow{2}{*}{Classifier Type} & \multicolumn{2}{c|}{Metrics} &\multirow{2}{*}{Classifier Classes} \\
        &Accuracy&AUROC&\\\midrule
        Age Group&0.74&0.69&[0,30], (30,40], (40,50], (50,60], (60,70], (70,80], (80,$\infty$)\\
        Race&0.76&0.64&Asian, Black or African American, White, Others\\
        Gender&0.92&0.87&Female, Male\\
        Tumor Location&0.89&0.73&Lung, Kidney, Colon and rectum\\
    \bottomrule
    \end{tabular}}
    \caption{\textbf{Classifier performance on the validation split of TCIA and RJ datasets}. Image-prompt alignment is assessed by evaluating the accuracy between features specified in input prompts and those predicted from generated images using these classifiers.}
    \label{tab:classifier}
\end{table}

\begin{table}[t]
    \centering
    \resizebox{\linewidth}{!}{
    \begin{tabular}{c|cccc|cccc}
    \toprule
       \multirow{2}{*}{Method}  & \multicolumn{4}{c|}{DSC$\uparrow$} & \multicolumn{4}{c}{HD$_{95}\downarrow$} \\
       &LU & CO & KI & Avg.&LU & CO & KI & Avg.\\\midrule
       GuideGen  & \textbf{0.71}& 0.21 &\textbf{0.64} &0.52 &\textbf{8.4} &227.0& 84.5 &106.6\\
       DiffTumor~\cite{chen2024towards} & 0.55 & 0.12&0.49&0.39&46.4&291.5&92.2&143.4\\
       LesionDiffusion~\cite{tian2025lesiondiffusion} & 0.69 & \textbf{0.36} &0.63&\textbf{0.56}&11.0&\textbf{128.3}&\textbf{79.1}&\textbf{72.8}\\\bottomrule
    \end{tabular}}
    \caption{\textbf{Performance of tumor segmentation model pretraining compared to tumor inpainting methods}. We report the DSC and HD$_{95}$ metrics of three tumor segmentation datasets as in the main paper.}
    \label{tab:difftumor}
\end{table}

\end{document}